**Directional vortex motion guided by artificially induced mesoscopic potentials.**

J. E. Villegas[1], E. M. Gonzalez[1], M. I. Montero[2], Ivan K. Schuller[2], and J. L. Vicent[1]

[1] Departamento de Física de Materiales, Facultad CC. Físicas, Universidad Complutense, 28040 Madrid (Spain).

[2] Department of Physics, University of California-San Diego, La Jolla, CA 92093 –0319 (USA).

**ABSTRACT.**

Rectangular pinning arrays of Ni dots define a potential landscape for vortex motion in Nb films. Magnetotransport experiments in which two in-plane orthogonal electrical currents are injected simultaneously allow selecting the direction and magnitude of the Lorentz force on the vortex-lattice, thus providing the angular dependence of the vortex motion.

The background dissipation depends on angle at low magnetic fields, which is progressively smeared out with increasing field. The periodic potential locks in the vortex motion along channeling directions. Because of this, vortex-lattice direction of motion is up to $85^{o}$ away from the applied Lorentz force direction.

Recently, e-beam writing based nanolithography[1] has been used to prepare submicrometric structures with geometries defined at will. In particular, the fabrication of superconducting thin films with periodic arrays of magnetic or non-magnetic dots[2,3], lines[4], or holes[5], with sizes comparable to characteristic lengths that govern superconductivity (the coherence length $\xi$ and the penetration depth $\lambda$), opened the door to studies of static and dynamic properties of vortex matter. One of the most remarkable phenomena observed in samples with periodic arrays of defects is the commensurability effect in magnetoresistance and critical current[2-14], which show that the vortex-lattice is strongly pinned at applied fields for which geometric matching exists between the vortex-lattice and the underlying periodic structure.

Vortex-lattice dynamics as a function of Lorentz force direction in this kind of samples has been investigated both theoretically[15,16] and experimentally[17] only for two privileged directions recently. In Nb thin films with rectangular arrays of magnetic dots[17], the vortex motion is easier when the Lorentz force is applied parallel to the short than the long side of the rectangular array. This effect was related to channeled potential barriers for vortex motion in the shortest inter-dot distance direction.

In the present work, we have studied vortex–lattice dynamics as a function of the Lorentz force direction with respect to the array axes, in samples with rectangular array symmetries. For this purpose, we have developed experiments in which the direction of the Lorentz force is rotated any angle at will. We have found that the vortex-lattice is guided by channeled potentials along the short side of the array even though Lorentz force is applied up to 85º away of the short side of the rectangular array.

Arrays of submicrometric Ni dots on Si (100) substrates were fabricated using e-beam lithography techniques. Briefly, the pattern is defined by e-beam writing on the resist covering the substrate, followed by developing, and Ni sputter deposition. After lift-off only the Ni nanostructures remain on top of the substrate, which is then covered by a sputtered Nb thin film. Further details on this procedure can be found elsewhere[1].

Two arrays with rectangular symmetry but different periods $a \times b$ were used; sample A with 400x625 nm$^2$ unit cell and sample B with 400 x 500 nm$^2$ unit cell. On both samples, the dots are 40 nm high (Ni thickness) and their diameter is $\varnothing$=250 nm. The Nb film on top of dot arrays is 100 nm thick.

For magnetotransport measurements, a cross-shaped bridge, designed for these rotating Lorentz force experiments, was optically lithographed and ion-etched. A sketch of the bridge, as well as notation and definition of angles and directions is shown in Figs. 1 (a) and (b). The two injected dc currents $J_x$ and $J_y$ cross in a square area containing the dot arrays. Taking into account Lorentz force expression, $\vec{F}_L = \vec{J} \times \vec{n} \phi_0$ (where $\phi_0 = 2.07 \cdot 10^{-15}$ Wb and $\vec{n}$ is a unitary vector parallel to the applied magnetic field), $J_x$ and $J_y$ yield two components of the Lorentz force on the vortex lattice, $F_x = J_y \phi_0$ and $F_y = J_x \phi_0$. Thus, the resultant Lorentz force, of magnitude $F_L = \sqrt{F_x^2 + F_y^2}$ and direction $\theta = arctag(J_y/J_x)$, is selected at will, and it can be rotated in-plane in the whole angular range 0-90 º, being parallel to the short lattice parameter $a$ when $\theta = 0º$ and to the long one $b$ for $\theta = 90º$. Voltage drops $V_x$=V$_2$-V$_3$ and $V_y$ = V$_1$-V$_2$ were simultaneously measured using two nanovoltmeters. Experiments were carried out in a liquid He cryostat, provided with a superconducting magnet. In all measurements, the magnetic field was always applied perpendicular to film plane, and thus perpendicular to the applied currents $J_x$ and $J_y$.

Normal state (T=9.5 K) voltage drops $V_x$ and $V_y$ as a function of $\theta = arctag(J_y/J_x)$ (Fig. 1 (c)) show sinusoidal behavior, proving that voltage contacts are well aligned. Superconducting critical temperatures $T_c$'s were independent of the measuring voltage contacts. For sample A it was $T_c$=8.75 K, and for sample B $T_c$=8.63 K, close to the $T_c$=9.2 K for bulk Nb.

The resistance along the $x$ direction (long side $b$ of rectangular array) $R_x = V_x/I_x$ versus applied field H is shown in Fig. 2 for different angles $\theta$ and fixed magnitude of the Lorentz force $F_L$, for samples A and B. For both samples, minima develop as a consequence of geometrical matching between vortex-lattice and the underlying periodic structure[2]. For sample A two different periods $\Delta H$ are observed: at low fields $\Delta H_{low}$=84 Oe, whilst at higher fields $\Delta H_{high}$=130 Oe. The low-field period corresponds to fields at which an integer number of vortices exists per unit cell of the array, giving $\Delta H_{low} = \phi_0/ab = 82.9$ Oe, where $a$=625 nm and $b$=400 nm are dots array cell dimensions. The high-field period corresponds to $\Delta H_{high} = \phi_0/a^2 = 129$ Oe. The transition between these two different regimes has been explained in terms of the reconfiguration in the vortex lattice from rectangular to square geometry[18]. For sample B ($a$=500 nm and $b$=400 nm), this phenomenon is also observed, with a longer period $\Delta H_{high}$=122 Oe than the low field regime $\Delta H_{low}$=104 Oe, in good agreement with theoretical values $\Delta H_{low} = \phi_0/ab = 103$ Oe and $\Delta H_{high} = \phi_0/a^2 = 129$ Oe. In addition, fractional matching effects[14] are observed for sample B, at fields $0.5\Delta H_{low}$=52 and $1.5\Delta H_{low}$=156 Oe. None of these features depend on the direction of the Lorentz force, i.e. the position and period of matching fields do not change when Lorentz force is rotated at different angles $\theta$. However, for both samples, the background resistance in the low field regime clearly depends on Lorentz force direction, and becomes lower as it is rotated towards the long side of the rectangular array $b$ ($\theta \rightarrow 90$). At higher fields, the resistance becomes the same for all angles $\theta$. This behavior clearly shows an anisotropic in-plane vortex dynamics.

Fig. 3 shows a different set of R(H) for samples A and B, with $R_x = V_x/I_x$ and $R_y = V_y/I_y$ measured simultaneously. In these measurements, the current density $J_x$ is kept constant, whilst $J_y$ is changed for each curve. As a result, the Lorentz force $F_y$ along the short side of the array $a$ is kept constant, and the force $F_x$ along the long side $b$ is increased, thus resulting in a total variable Lorentz force $F_L = \sqrt{F_x^2 + F_y^2}$ with direction $\theta = arctag(J_y/J_x)$. The behavior of both A and B samples is essentially the same (Fig. 3). For fields up to the first matching and $\theta < 85°$, $R_x$ is independent of $F_x$. Moreover, in these field range, $R_y$ falls below measurable values, much lower than $R_x$; that is $V_y << V_x$, indicating that vortex lattice is essentially moving along the short side $a$ of the array. It is worth noting that vortex-lattice motion is confined along the short side $a$ of the array, even though the total Lorentz force $F_L$ is rotated up to $\theta=85°$, very close to the direction of the long side of the array $b$. For fields higher than the first matching, a measurable resistance $R_y$ progressively arises. Thus, the vortex-lattice direction of motion no longer is restricted to the $a$ direction, but it rotates as magnetic field increases, becoming parallel to the direction of the applied Lorentz force $F_L$.

The phenomenology described above constitutes the central result of this paper. The anisotropic pinning potential due to the rectangular array of magnetic dots[17] creates a hard-axis, along the short side $b$, for vortex-lattice motion. The important point is that this effect is strong enough to lock vortex motion along the short side $a$ of the rectangle. Vortex-lattice motion is guided along this direction for fields up to the first matching field, disappearing progressively for more intense fields; as the number of weakly pinned interstitial vortices[2] increases the guided vortices effect is reduced.

To gain further insight into this anisotropic vortex-lattice dynamics, I-V curves (not shown) were measured at T=0.99T$_c$ for several $\theta = arctag(J_y/J_x)$ at different applied fields. The vortex-lattice velocities along the $x$ (hard) and $y$ (easy) axes are calculated from the voltage drops $V_x$ and $V_y$, using $v_i = V_i/(dB)$, where $d$ is the distance between contacts and $B$ the applied field. Here we used that the vortex-lattice velocity $v = \sqrt{v_x^2 + v_y^2}$ gives the electric field $\vec{E} = \vec{B} \times \vec{v}$. Together with the Lorentz force expression, $F_L = \sqrt{F_x^2 + F_y^2}$, we can derive $F_L$-$v$ curves from I-V characteristics. This is interesting since[19] a comparison of the Lorentz force necessary to drive the vortex-lattice with a velocity $v$ at out-of-matching field ($F_{L\,out}$) with that at matching ($F_{L\,matching}$), gives the effective force at matching ($\Delta F_L$) as a function of velocity,

$$F_{L\,matching}(v) - F_{L\,out}(v) \equiv \Delta F_L(v)$$

This analysis for the first matching field is shown in Fig. 4, for sample A ($H_{out}$=45 Oe, $H_{matching}$=84 Oe) and B ($H_{out}$=70 Oe, $H_{matching}$=104). Figs. 4 (a) and (b) show that, for angles up to $\theta = 60°$, the maxima of the effective force are for velocities around 200 m/s in both samples, well in the range earlier reported[19], and, most important, do not depend on the direction of the Lorentz force. However, as Lorentz force direction approaches the hard-axis $b$ ($60 < \theta \to 90$, perpendicular to the channel direction), the maxima shift towards lower velocities. On the other hand, $\Delta F_L(\theta)$ increases monotonically for all $\theta$.

Figs. 4 (c) and (d) show $\Delta F_L(\theta) \times \cos(\theta)$. $\Delta F_L(\theta) \times \cos(\theta) \approx \Delta F_L(\theta = 0)$ up to a given velocity, which depends on sample anisotropy and Lorentz force direction ($\theta$). This clarifies the behavior observed in Figs. 4 (a) and (b). Up to a given vortex-lattice velocity, dependent on Lorentz force direction, only the $F_y$ component (along easy-axis $a$) plays a role in vortex dynamics. In agreement with the picture described above, the vortex motion is restricted to this direction for fields up to first matching field. The component of the Lorentz force along the hard-axis $b$, $F_x$, has no effect until a threshold for vortex-lattice velocity is reached. For both samples, this threshold is lower as the direction of the Lorentz force is closer to the hard-axis (see arrows in Figs. 4 (a) and (b)). The more anisotropic is the rectangular array (sample A), the higher the threshold velocity for each Lorentz force direction.

In conclusion, measurement of the vortex-lattice velocity $v$ versus applied Lorentz force $F_L$ shows that, up to a given vortex-lattice velocity threshold, the vortex motion in rectangular pinning arrays is strongly guided by the microstructure along the short side of the array. The physical origin of this guided vortex motion is the channeling potential landscape created by the rectangular symmetry. This results in vortex-lattice velocity and Lorentz force not being parallel for fields up to the first matching minimum: the effect is strong enough to keep vortex-lattice motion up to 85º away from the Lorentz force direction., and only the component of the Lorentz force $F_y$ along the short side of the array plays a role in vortex dynamics. The effect is tunable by the dots array anisotropy: the higher the anisotropy of the dot array, the wider the range of velocities in which the guided vortex motion occurs.

We acknowledge grant MAT02-04543 from Spanish CICYT, Fundación Ramón Areces, New Del Amo Program and ESF-VORTEX Program. Work at UCSD supported by the US-DOE. We thank Prof. J. Santamaría for useful conversations and critical reading of the manuscript, and thank A. Silhanek and J. Schuller for helpful conversations.

[1] J.I. Martín, J. Nogués, K. Liu, J.L. Vicent and I. K. Schuller, J. Magn. Magn. Mat. **256**, 449 (2002).

[2] J.I. Martín, M. Vélez, J. Nogués, and Ivan K. Schuller, Phys. Rev. Lett. **78**, 1929 (1997)

[3] Y. Jaccard, J. I. Martín, M.-C. Cyrille, M. Vélez, J. L. Vicent, and I. K. Schuller Phys. Rev. B **58**, 8232 (1998).

[4] D. Jaque, E. M. González, J. I. Martín, J. V. Anguita, and J. L. Vicent, Appl. Phys. Lett. **81**, 2851 (2002).

[5] L. Van Look, B. Y. Zhu, R. Jonckheere, B. R. Zhao, Z. X. Zhao, and V. V. Moshchalkov, Phys. Rev. B **66**, 214511 (2002).

[6] O. Daldini, P. Martinoli, J.L. Olsen, and G. Berner, Phys. Rev. Lett. **32**, 218 (1974).

[7] A.T. Fiory, A.F. Hebard, and S. Somekh, Appl. Phys. Lett. **32**, 73 (1978).

[8] A. Pruymboom, P.H. Kes, E. van der Drift, and S. Radelaar, Phys. Rev. Lett. **60**, 1430 (1988).

[9] Y. Otani, B. Pannetier, J.P. Nozières, and D. Givord, J. Magn. Magn. Mater. **126**, 622 (1993).

[10] M. Baert, V. Metlushko, R. Jonckheere, V.V. Moshchalkov, and Y.Bruynseraede, Phys. Rev. Lett. **74**, 3269 (1995).

[11] D.J. Morgan and J.B. Ketterson, Phys. Rev. Lett. **80**, 3614 (1998).

[12] V. Metlushko, U. Welp, G.W. Crabtree, R. Osgood, S.D. Bader, L.E. De Long, Z. Zhang, S.R.J. Brueck, B. Ilic, K. Chung, and P.Hesketh, Phys. Rev. B **60**, R12585 (1999).

[13] Y. Fasano, J.A. Herbsommer, F. de la Cruz, F. Pardo, P. Gammel, E. Bucher, and D. Bishop, Phys. Rev. B **60**, R15047 (1999).

[14] O. M. Stoll, M. I. Montero, J. Guimpel, J. J. Åkerman, and I. K. Schuller, Phys. Rev. B **65**, 104518 (2002).

[15] C. Reichhardt and F. Nori, Phys. Rev. Lett**. 82**, 414 (1999).

[16] C. Reichhardt, G.T. Zimanyi, and N. Gronbech-Jensen, Phys.Rev. B **64**, 014501 (2001).

[17] M. Velez, D. Jaque, J. I. Martín, M. I. Montero, I. K. Schuller, and J. L. Vicent, Phys. Rev. B **65**, 104511 (2002).

[18] J. I. Martín, M. Vélez, A. Hoffmann, I. K. Schuller, and J. L. Vicent, Phys. Rev. Lett. **83**, 1022 (1999).

[19] M. Vélez, D. Jaque, J. I. Martín, F. Guinea, and J. L. Vicent, Phys. Rev. B **65**, 094509 (2002).

**FIGURE CAPTIONS**

**Figure 1:** (a) Micrograph of the measurement bridge. The area in which currents cross is 40x40 µm$^2$ . Shaded area represents the 90x90 µm$^2$ array of dots. (b) Definition of angle $\theta$, and current and voltage directions with respect to array axes. (c) Voltage drops $V_y=V_1-V_2$ and $V_x=V_2-V_3$ as a function of $\theta$ above Tc (T=9.5 K), with $\sqrt{J_x^2 + J_y^2}$ =10 kA cm$^{-2}$ (d) SEM image of the array of sample A (*a*=400 nm, *b*=625 nm).

**Figure 2:** (a) $R_x$ *(H)* vs. applied field of sample A for different angles $\theta$ and $\sqrt{J_x^2 + J_y^2}$ =75 kA cm$^{-2}$ at T=0.98T$_c$ . Vertical dashed lines point out matching fields with period $\Delta H_{low}$=84 Oe, and dashed-dot lines those with $\Delta H_{high}$=130 Oe. (b) $R_x$ *(H)* vs. applied field of sample B for different angles $\theta$ and $\sqrt{J_x^2 + J_y^2}$ =12.5 kA cm$^{-2}$ at T=0.99T$_c$. Vertical dashed lines point out matching fields with period $\Delta H_{low}$=104 Oe, dashed-dotted lines those with $\Delta H_{high}$=122 Oe, and dotted lines mark fractional matching fields 0.5$\Delta H_{low}$=52 Oe and 1.5$\Delta H_{low}$=156 Oe.

**Figure 3:** (a) $R_x$ *(H)* and (b) $R_y$*(H)* of sample A at T=0.99T$_c$, for $J_x$=25 kA cm$^{-2}$ and different $J_y$: (solid squares) $J_y$=0 ($\theta$=0), (circles) $J_y$=12.5 kA cm$^{-2}$ ($\theta$=30), (solid up triangles) $J_y$ =43.3 kA cm$^{-2}$ ($\theta$=60) and (down triangles) $J_y$=68.8 kA cm$^{-2}$ ($\theta$=75). (c) $R_x$ *(H)* and (d) $R_y$*(H)* of sample B at T=0.99T$_c$, for $J_x$=12.5 kA cm$^{-2}$ and different $J_y$: : (solid squares) $J_y$=0 ($\theta$=0), (circles) $J_y$=7.25 kA cm$^{-2}$ ($\theta$=30), (solid up triangles) $J_y$=12.5 kA cm$^{-2}$ ($\theta$=45), (up triangles) $J_y$=21.6 kA cm$^{-2}$ ($\theta$=60), (solid diamonds) $J_y$=34.3 kA cm$^{-2}$ ($\theta$=70) and (left triangles) $J_y$=142.2 kA cm$^{-2}$ ($\theta$=85). Dashed line points out first matching fields.

**Figure 4:** (a) $\Delta F_L(v)$ for sample B around first matching, for different directions of the Lorentz force (see legend). (b) $\Delta F_L(v)$ for sample A around first matching, for different directions of the Lorentz force (see legend). (c) $\Delta F_L(v) \times \cos(\theta)$ for sample B around first matching (see legend). (d) $\Delta F_L(v) \times \cos(\theta)$ for sample A around first matching (see legend). Horizontal arrows in (c) and (d) point out vortex velocity threshold for guided vortex motion.

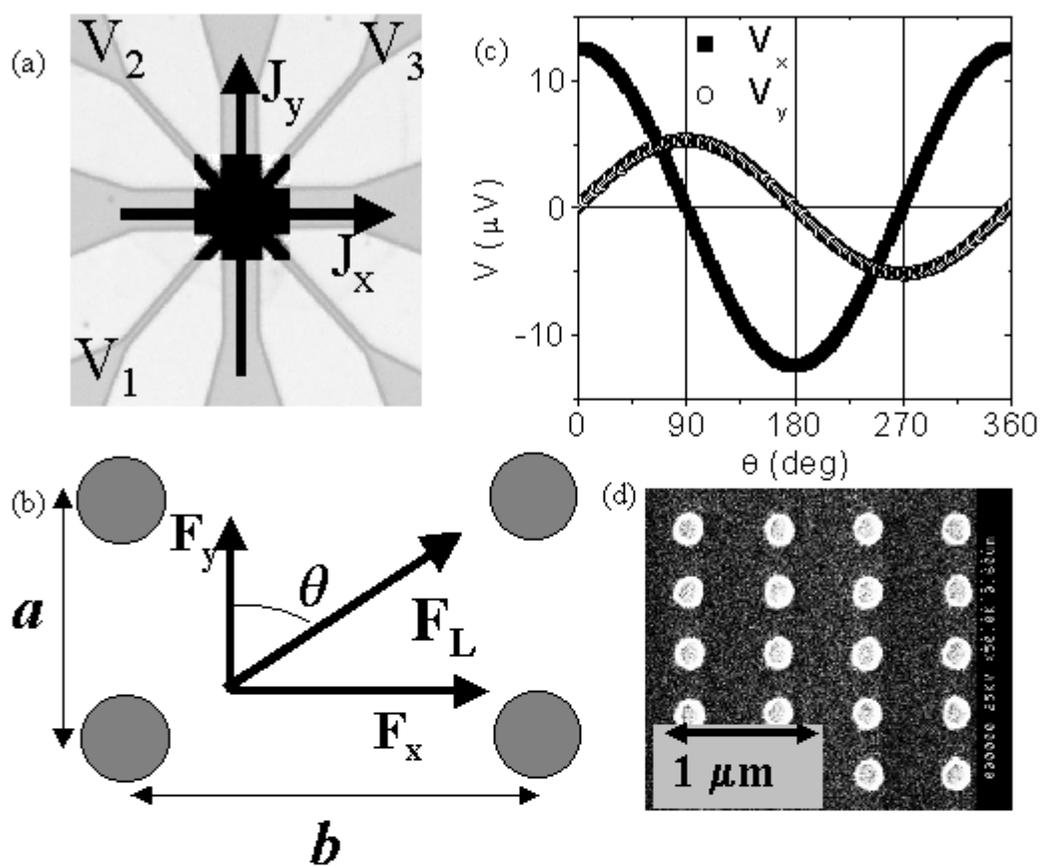

Figure 1
J.E. Villegas *et al.*

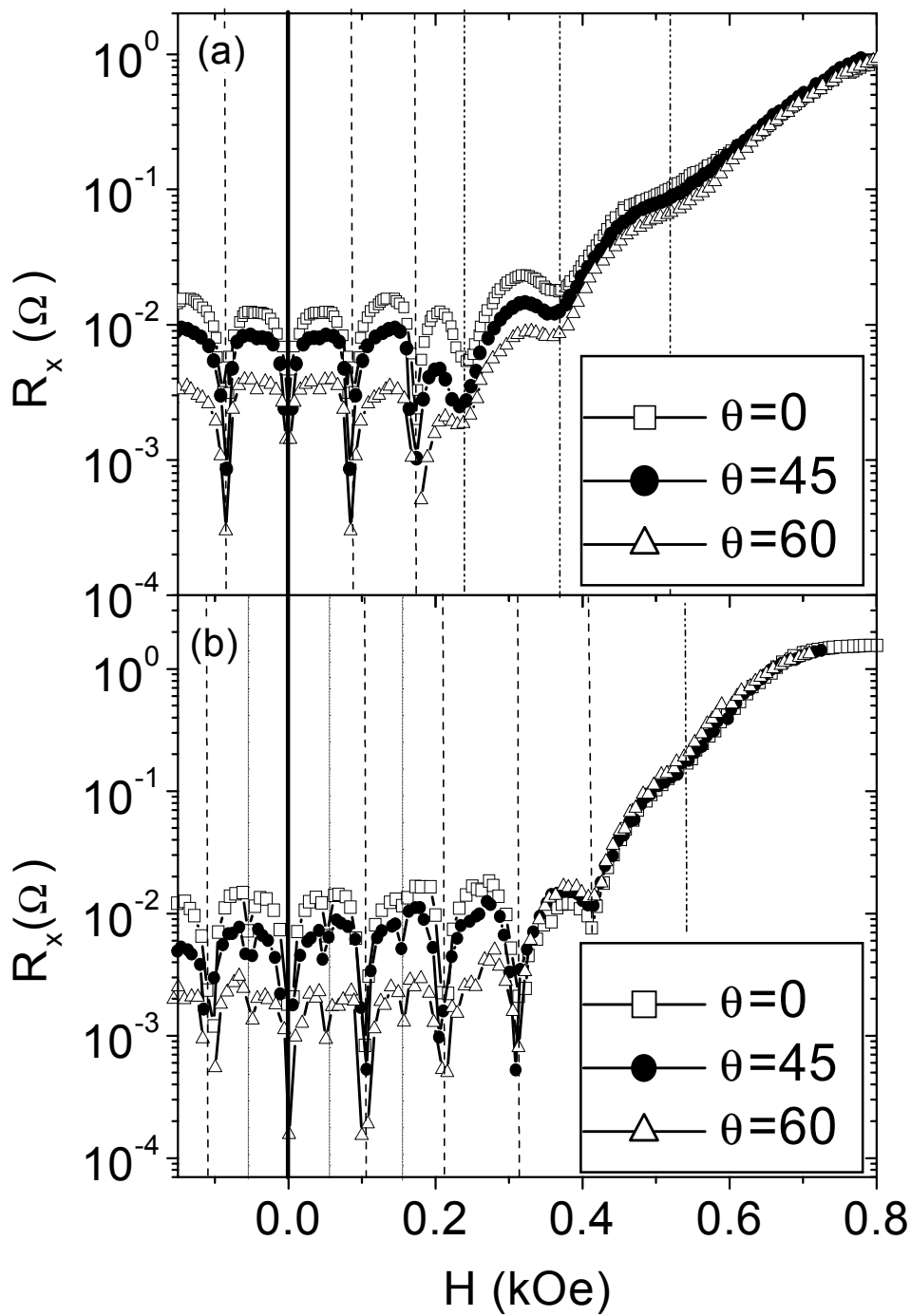

Figure 2
J.E. Villegas *et al.*

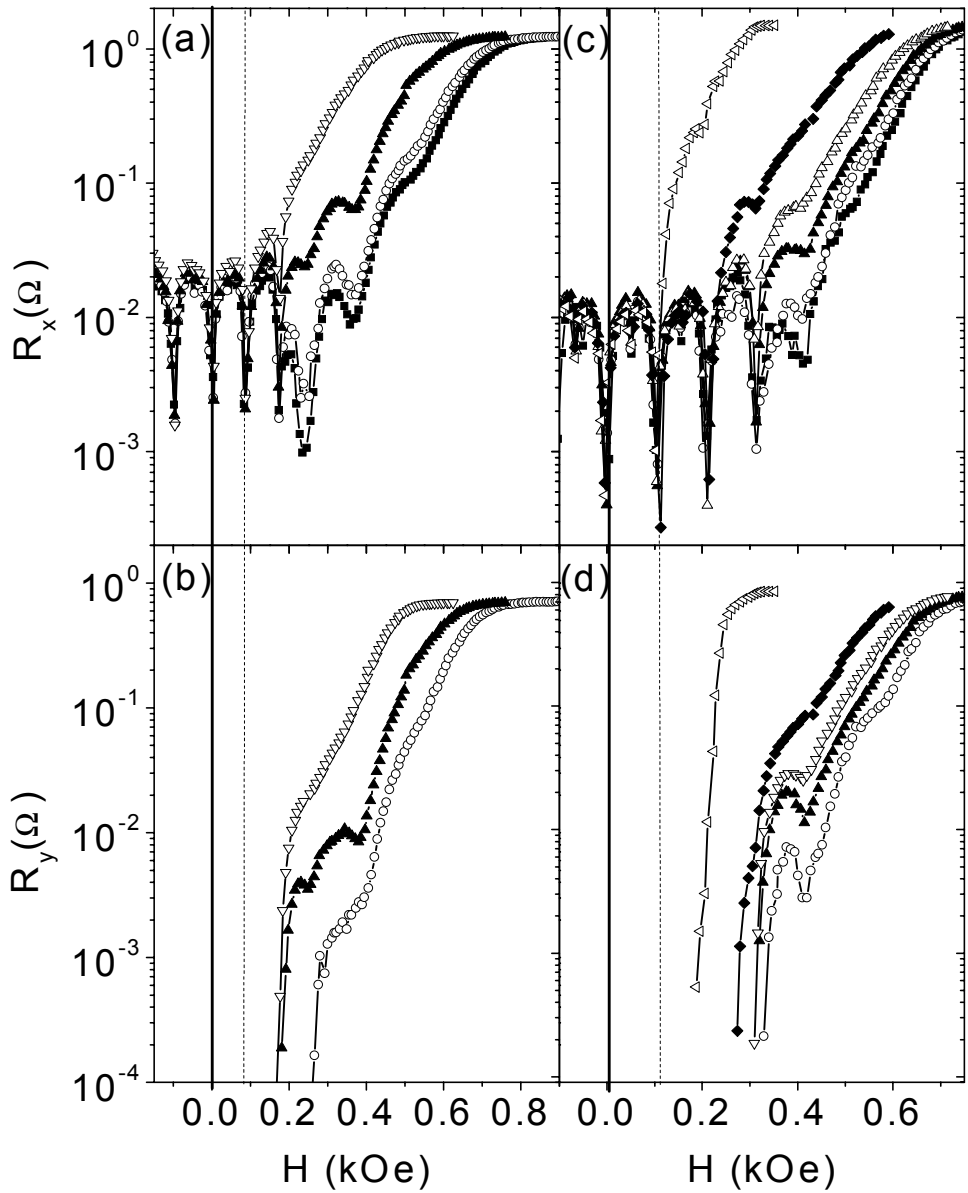

Figure 3
J.E. Villegas *et al.*

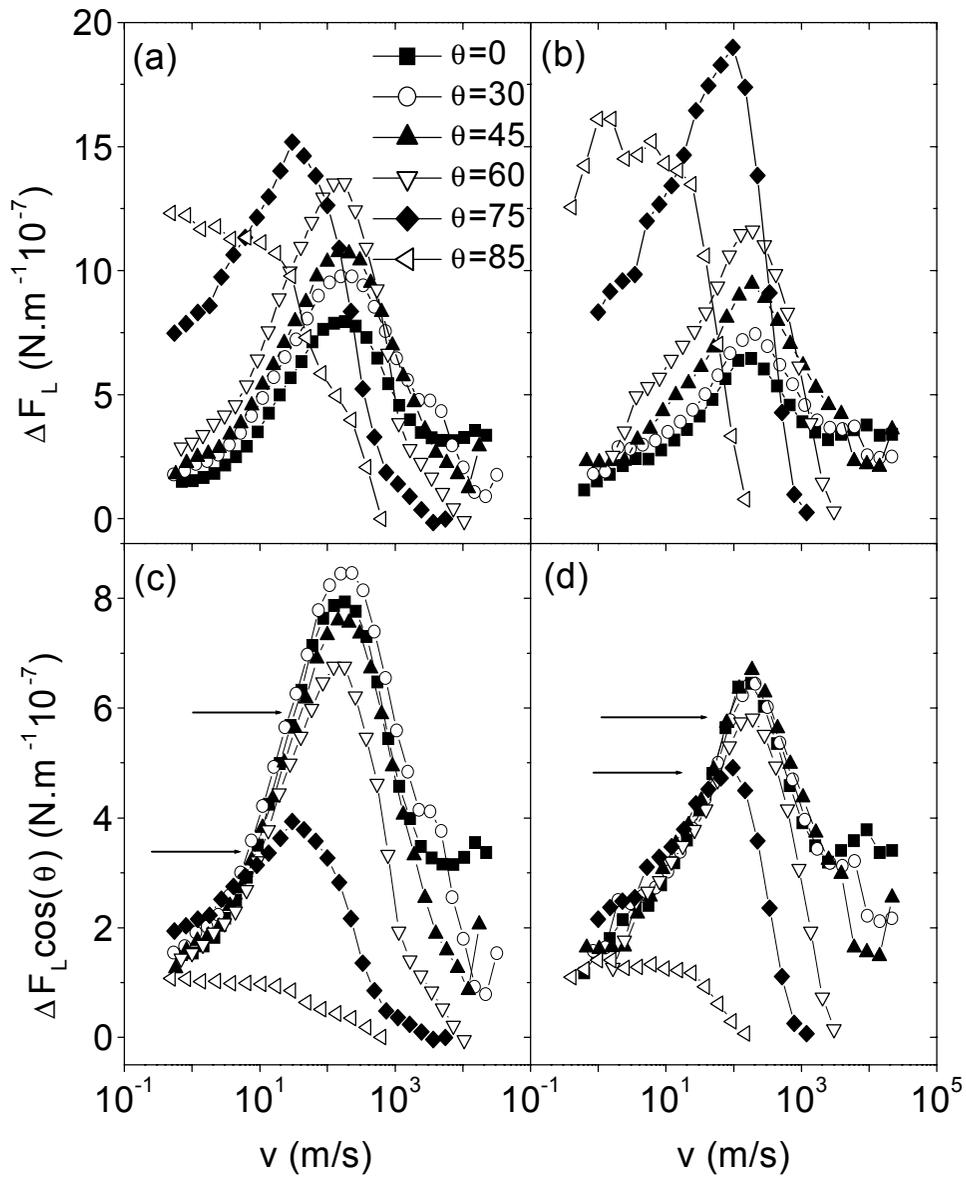

Figure 4
J.E. Villegas *et al.*